\documentclass[sigconf,screen]{acmart}
\AtBeginDocument{%
  \providecommand\BibTeX{{%
    \normalfont B\kern-0.5em{\scshape i\kern-0.25em b}\kern-0.8em\TeX}}}

\setcopyright{acmlicensed}
\acmPrice{15.00}
\acmDOI{10.1145/3611643.3613880}
\acmYear{2023}
\copyrightyear{2023}
\acmSubmissionID{fse23industry-p61-p}
\acmISBN{979-8-4007-0327-0/23/12}
\acmConference[ESEC/FSE '23]{Proceedings of the 31st ACM Joint European Software Engineering Conference and Symposium on the Foundations of Software Engineering}{December 3--9, 2023}{San Francisco, CA, USA}
\acmBooktitle{Proceedings of the 31st ACM Joint European Software Engineering Conference and Symposium on the Foundations of Software Engineering (ESEC/FSE '23), December 3--9, 2023, San Francisco, CA, USA}
\received{2023-05-18}
\received[accepted]{2023-07-31}

\usepackage{amsmath,amsfonts}
\usepackage{algorithmic}

\usepackage{graphicx}
\usepackage{textcomp}
\usepackage{xcolor}
\usepackage{tcolorbox}
\usepackage{graphicx}
\usepackage{listings}
\usepackage{makecell}
\usepackage{adjustbox}
\usepackage{multirow}
\usepackage{graphicx}
\usepackage{svg}
\usepackage{xspace}
\usepackage{amsmath}
\usepackage{subfigure}
\usepackage{caption}
\usepackage{amsmath}
\usepackage{booktabs}
\usepackage{color, colortbl}
\usepackage{url}
\usepackage{caption} 
\usepackage{pifont}
\usepackage{tabularx}

\newcolumntype{P}[1]{>{\centering\arraybackslash}p{#1}}
\usepackage[ruled,vlined,linesnumbered]{algorithm2e}

\SetCommentSty{mycommfont}
\usepackage{pifont}
\usepackage{enumitem}

\usepackage{caption}
\newcommand{\xmark}{\ding{55}}
\newcommand{\cmark}{\ding{51}}
\newcommand{\tool}{{\texttt{WeReplay}}\xspace}

\usepackage{xcolor}
\usepackage{xcolor,colortbl}
\definecolor{lightgray}{gray}{0.93}
\definecolor{slightgray}{gray}{0.98}
\definecolor{darkgray}{gray}{0.77}

\definecolor{amber}{rgb}{1.0, 0.49, 0.0}

\usepackage{framed}
\definecolor{formalshade}{rgb}{0.95, 0.95, 1}
\definecolor{mygray}{gray}{0.4}
\definecolor{lightgray}{gray}{0.93}

\begin{document}

\title{Towards Efficient Record and Replay: A Case Study in WeChat}

\author{Sidong Feng}
\affiliation{%
  \institution{Monash University}
  \city{Melbourne}
  \country{Australia}}
\email{sidong.feng@monash.edu}

\author{Haochuan Lu}
\affiliation{%
  \institution{Tencent Inc.}
  \city{Guangzhou}
  \country{China}}
\email{hudsonhclu@tencent.com}

\author{Ting Xiong}
\affiliation{%
  \institution{Tencent Inc.}
  \city{Guangzhou}
  \country{China}}
\email{candyxiong@tencent.com}

\author{Yuetang Deng}
\affiliation{%
  \institution{Tencent Inc.}
  \city{Guangzhou}
  \country{China}}
\email{yuetangdeng@tencent.com}

\author{Chunyang Chen}
\affiliation{%
  \institution{Monash University}
  \city{Melbourne}
  \country{Australia}}
\email{chunyang.chen@monash.edu}

\renewcommand{\shortauthors}{Feng et al.}

\begin{abstract}
  WeChat, a widely-used messenger app boasting over 1 billion monthly active users, requires effective app quality assurance for its complex features. Record-and-replay tools are crucial in achieving this goal. Despite the extensive development of these tools, the impact of waiting time between replay events has been largely overlooked. On one hand, a long waiting time for executing replay events on fully-rendered GUIs slows down the process. On the other hand, a short waiting time can lead to events executing on partially-rendered GUIs, negatively affecting replay effectiveness. An optimal waiting time should strike a balance between effectiveness and efficiency. We introduce \tool, a lightweight image-based approach that dynamically adjusts inter-event time based on the GUI rendering state. Given the real-time streaming on the GUI, \tool employs a deep learning model to infer the rendering state and synchronize with the replaying tool, scheduling the next event when the GUI is fully rendered. Our evaluation shows that our model achieves 92.1\% precision and 93.3\% recall in discerning GUI rendering states in the WeChat app. Through assessing the performance in replaying 23 common WeChat usage scenarios, \tool successfully replays all scenarios on the same and different devices more efficiently than the state-of-the-practice baselines.
\end{abstract}

\begin{CCSXML}
<ccs2012>
   <concept>
       <concept_id>10011007.10011074.10011099.10011102.10011103</concept_id>
       <concept_desc>Software and its engineering~Software testing and debugging</concept_desc>
       <concept_significance>500</concept_significance>
       </concept>
 </ccs2012>
\end{CCSXML}

\ccsdesc[500]{Software and its engineering~Software testing and debugging}

\keywords{Efficient record and replay, GUI rendering, Machine Learning}

\maketitle

\section{Introduction}
WeChat\footnote{\url{https://www.wechat.com/}}, with over 1.67 billion monthly active users, is among the most popular messenger apps in the world, particularly for those of Chinese origin~\cite{web:wechat_statistic}.
In fact, WeChat has now evolved beyond a simple messaging app, now offering features such as banking, shopping, gaming, news browsing, and serving as a platform for third-party app development~\cite{wang2022characterizing,gao2019emerging}.
As WeChat rapidly expands its features, effective app quality assurance becomes increasingly crucial and challenging.
Record-and-replay tools have long been a vital approach, enabling developers to record app usage scenarios and automate their replay on various devices.
Numerous record-and-replay tools have been developed by both practitioners and researchers to address this need~\cite{zadgaonkar2013robotium,gomez2013reran,web:culebra,web:hiroMacro,web:monkeyrunner,web:espresso,halpern2015mosaic,hu2015versatile,web:appetizer,qian2020roscript,long2020webrr,li2022cross}.
% Typically, these tools timely record the user events and automating replay them in the same manner across different devices, i.e., replaying the same event after a same time interval.

\begin{figure*}
	\centering
	\includegraphics[width=0.85\linewidth]{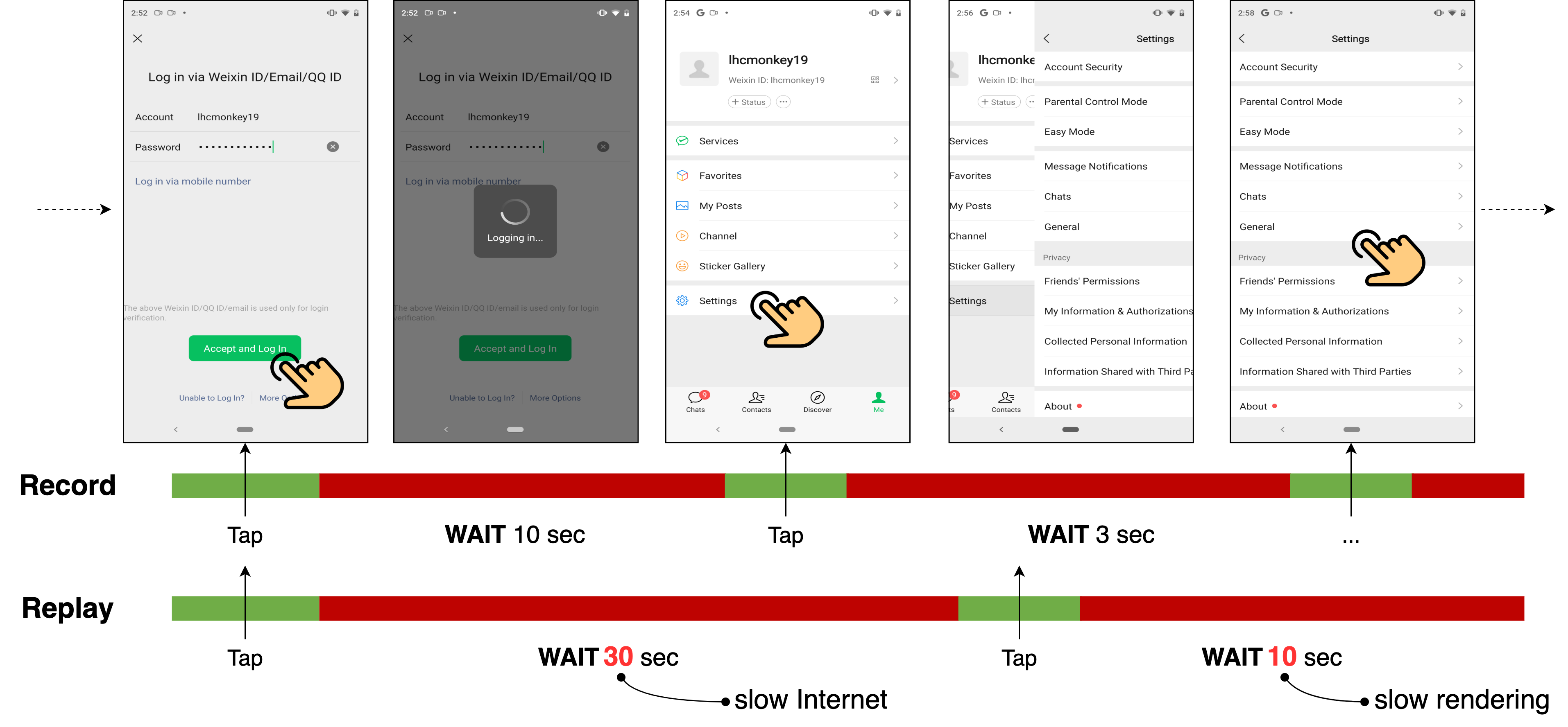}
	\caption{Example of record and replay events. Green bars represent event triggers, while red bars represent the inter-event waiting time. A fixed waiting time captured during the recording may not be sufficient for replaying, as dynamic factors such as slow internet or delayed rendering can affect the process.}
	\label{fig:flow}
\end{figure*}

While these tools perform reliably in many apps, there remains a significant gap in industrial capability. 
One often overlooked aspect is the waiting time between replaying events. 
Typically, record-and-replay tools record the time delay between events and automatically replay them on devices with the same elapsed time. 
However, our study of nine WeChat usage scenarios reveals that a fixed time elapse may not be practical due to two main reasons. 
First, GUI loading for replaying may take longer due to Internet-related factors, which are common in the WeChat app such as logging in, sending messages, etc. 
Second, replaying on older devices may require more time due to decreased rendering efficiency and lower processing capability. 
These issues of executing replay events on incompletely rendered GUIs impede successful replay on the same device (with a 45\% failure rate) and different devices (with a 63\% failure rate).

To ensure effective replaying, WeChat developers usually manually set event time delays to a considerably longer waiting time — about 10 times longer than the recorded duration. 
However, this manual setting process can be time-consuming and error-prone for developers, taking an average of 10 minutes to set up 5 events, according to the WeChat developers. 
Moreover, replay becomes inefficient with idle waiting on fully rendered GUIs, slowing down the testing process. 
Due to budget constraints and market pressures in the industry, development teams often face limited testing time. 
Consequently, the more efficient the replay tools, the more devices can be tested, increasing the likelihood of discovering bugs and ultimately improving the quality of the app release.

Towards that target, we draw inspiration from previous work~\cite{feng2023efficiency} to accelerate automated app testing with GUI rendering inference.
While the app under testing is mostly ideal, the replay tool has to wait until the GUI finishes rendering before moving to the next event. 
With this in mind, we introduce \tool, a lightweight record-and-replay tool designed to dynamically schedule replay events by distinguishing between fully rendered and partially rendered GUIs.
To record the events to WeChat, we leverage the app inspector tool WEditor to retrieve the widget information and parse it to the testing script.
To replay the events in the testing script dynamically, we adopt a deep learning method to model the visual information from the GUI screenshot for inferring the GUI state.
Specifically, we first develop a tailored method to collect a large-scale binary GUI dataset from WeChat, comprising 4,485 fully rendered and 6,171 partially rendered GUIs. 
Next, we adopt a small but efficient Convolutional Neural Network (CNN) based approach to discern the GUI rendering state.
To implement our approach, ensuring events are sent only when the GUI is fully rendered, we establish a synchronization framework that streams real-time GUI screenshots and schedules replay events based on GUI rendering inference. 
Notably, one advantage of our approach is its purely image-based nature, making it easily deployable in other industrial apps.

% Record-and-replay tools has long been an important approach, allowing developers to record usage scenarios with apps and replay them on different devices to ensure the quality of apps before their delivery to end users.
% However, due to the budget limit and market pressure in the industry, development teams have to meet deadlines where the testing time is limited.
% Therefore, the less replay time of the tools, the more devices can be tested, the more likely to find bugs, and the higher quality of the app release.
% While much effort from academia and the industry has been dedicated in building record-and-replay tools~\cite{??}, a few of them focus on the replay efficiency by leveraging advanced infrastructure support to execute events efficiently~\cite{??}.

To evaluate the accuracy of our approach, we carry out an experiment on the GUI screenshots from the WeChat app. 
Our approach outperforms nine state-of-the-art baselines and two ablation baselines in predicting GUI rendering states, achieving 92.1\% precision, 93.3\% recall, and 92.7\% F1-score, respectively.
We also carry out an experiment to demonstrate that our tool can accelerate the replaying process without sacrificing effectiveness by replaying 23 usage scenarios from the WeChat app. 
Given a recorded usage scenario from one device, our tool can accurately replay it on both the same device and different devices in less time, compared to the baselines and industrial solutions.

The contributions of this paper are as follows:
\begin{itemize}
    \item We propose \tool, a practical record-and-replay tool designed to adaptively adjust the waiting time between replaying events, for speeding up the replaying process.
    \item We conduct a motivational empirical study to understand the prevalence of waiting time issues in record-and-replay for WeChat, which serves as the foundation for our research.
    \item We perform comprehensive experiments, including an evaluation of \tool's performance and its industrial record-and-replay practice in WeChat, to demonstrate the accuracy, efficiency, and effectiveness of our tool.
\end{itemize}

\section{MOTIVATIONAL Study}
\label{sec:motivation}
% Many researches have been done to deal with the record and replay of test scripts. 
% To better understand the issues of record and replay in industrial apps, we carried out a pilot study to examine the prevalence of these issues, so as to facilitate the development of our tool to enhance the existing record-and-replay tools.

Many record-and-replay tools have been developed by both practitioners and researchers to assist developers in easily recording and replaying complex usage scenarios of apps.
To assess the practicality of these record-and-replay tools in industrial apps, we conducted a pilot study by applying state-of-the-practice tools to the most popular social media app, WeChat.

\subsection{Experimental Setup}
We collected an experimental dataset of nine common usage scenarios from WeChat to evaluate whether the record-and-replay tools can accurately record and replay such scenarios.
These scenarios were identified by WeChat developers and through our hands-on experiences using the app, with an average of 6.4 events per scenario. 
An example of a common usage scenario, ``opening setting page in WeChat'' is illustrated in Figure~\ref{fig:flow}, including inputting a username and password to log in to the app and accessing settings on the personal page.
% its description on the Google Play store 
% They are also verified by the internal WeChat's developers.
% All of the scenarios in which we evaluate the tools on are described in greater detail and shown through videos on our project website [13].

To record and replay these common usage scenarios, we chose the state-of-the-practice tool SARA~\cite{guo2019sara} for two reasons.
First, SARA is widget-sensitive, meaning that the recorded events are based on the attributes of the widget (e.g., resource id). 
This approach is more robust for replaying on multiple devices with different resolutions and GUIs compared to recording coordinates~\cite{yan2018land}. 
Second, SARA's recording is timing-sensitive, as it automatically captures the time between events and utilizes this information during the replay process. 
We did not adopt other publicly available record-and-replay tools (e.g., Robotium~\cite{zadgaonkar2013robotium}, Culebra~\cite{web:culebra}, etc.) due to their impracticality in industrial apps or the requirement for costly investments~\cite{lam2017record}.

\renewcommand{\arraystretch}{1}
\begin{table}
    \small
    \tabcolsep=0.06cm
	\centering
	\caption{Overview of devices. }
	\label{tab:device}
	\begin{tabular}{l|c|c|c|c} 
	    \hline
	    \bf{Device} & \bf{Resolution} & \bf{Size (inch)} & \bf{OS Version} & \bf{CPU Processor}\\
	      \hline
            \hline
           Xiaomi Mix2S & 1080x2160 & 5.99 & 9.0 & Snapdragon 845 \\
           Huawei Nova2S & 1080x1920 & 5.50 & 8.0 & Kirin 659 \\
           Vivo Y3 & 720x1544 & 6.35 & 9.0 & Helio P35   \\
           Google Pixel4a & 1080x2340 & 5.81 & 11.0 & Snapdragon 730G \\
            \hline
	\end{tabular}
\end{table}

\begin{figure*}
	\centering
	\includegraphics[width=0.99\linewidth]{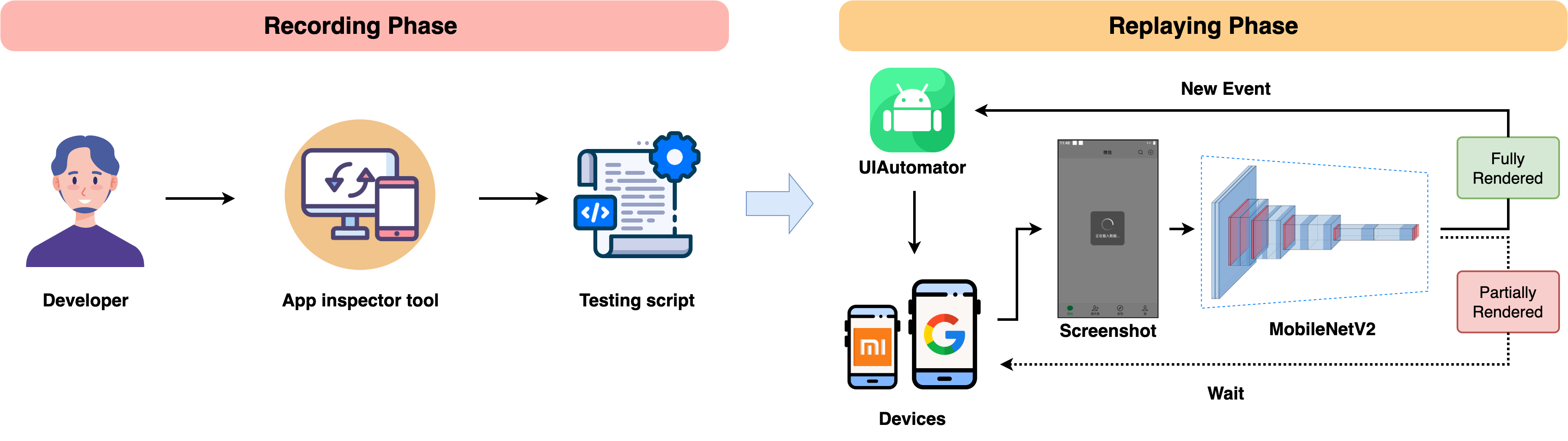}
	\caption{The workflow of our approach.}
	\label{fig:overview}
\end{figure*}

\subsection{Record and Replay on Same Device}
To understand the practice of record-and-replay in WeChat, we initially conducted a small pilot study focused on replaying usage scenarios recorded by SARA on the same device.
We utilized a Xiaomi Mix2S running Android 9.0 as the testing device, which is widely used in previous studies~\cite{yu2021layout}.
In total, we obtained 9 usage scenarios replays captured using screencast.
Two authors manually examined the states in the video replays to evaluate the correctness of each replay.

55\% of the scenarios were successfully replayed. 
We manually examined the failure cases and identified one common cause. 
The waiting time between events is dynamically affected by internet bandwidth. 
For example, consider a scenario where an app is logged and recorded in an environment with excellent internet connectivity, resulting in a shorter waiting time (i.e., 10 seconds in Figure~\ref{fig:flow}).
However, when replaying the scenario in an environment with poor internet connectivity, the replay fails due to the increased waiting time required to trigger the next event (i.e., 30 seconds in Figure~\ref{fig:flow}). 
WeChat developers further confirmed this phenomenon in recording and replaying industrial apps like WeChat, which frequently rely on internet connectivity.

\subsection{Record and Replay on Different Devices}
To evaluate the reproducibility of scenarios across different devices, we first recorded the scenarios using a Xiaomi Mix2S. 
We then replayed the events on three devices to ensure usability for a wide variety of users in practice. 
Table~\ref{tab:device} provides details of these devices, covering different resolutions, sizes, operating system versions, and CPU processors. 
These devices are widely used in previous studies for record-and-replay experiments~\cite{yu2021layout,romano2021empirical}.
In total, we obtained 27 replays and manually assessed the reproducibility of each one.

We discovered that 11\%, 33\%, and 66\% of the scenarios were successfully replayed for the Huawei Nova2S, Vivo Y3, and Google Pixel4a, respectively.
The former two devices replayed fewer scenarios than the latter, particularly the Huawei Nova2S, which could only replay 11\% of the scenarios.
The main reason is that a fixed amount of waiting time between events recorded on a more advanced device may not be suitable for replaying on older devices.
This can be attributed to factors such as decreased rendering efficiency and lower processing capability often associated with older devices.
WeChat developers further confirmed that unexpected device lagging can significantly impact the performance of scenario replays.

\subsection{Industrial Solution \& Motivation}
WeChat developers confirm the significance of the waiting time settings for recording and replaying on both the same and different devices. 
They also emphasize that setting arbitrary time delays between events is undesirable, as they are likely to fail due to insufficient loading time between events. 
To ensure that the recorded scenario is 100\% replayable on any device, WeChat developers need to manually set the time delay of events to a relatively long waiting time, usually 10x longer than the recorded waiting time, as shown in Table~\ref{tab:waiting}. 
However, this approach raises two practical issues. First, the manual setting can be time-consuming, taking more than 10 minutes on average for a 5-event scenario, according to WeChat developers. 
Second, the replay process is inefficient, frequently stagnating on fully loaded GUIs and slowing down the replaying process, i.e., taking over 3.5 minutes on average to replay 5 events.

% These findings confirm the importance of waiting time setting to record and replay to same devices and different devices.
% However, setting an optimal waiting time between events is not a trivial task. 
% On the one hand, setting a short waiting time will cause the events to execute on partially loaded GUIs, which negatively affects the replaying effectiveness. 
% Therefore, the developers of WeChat typically set a long waiting time to execute events to guarantee the robustness of replaying.
% However, it may inefficiently stagnate on fully loaded GUIs, slowing down the replaying process.
While the app under testing is mostly ideal, the replaying has to wait until the GUI finishes rendering before proceeding to the next event. 
In this context, it is worth developing an effective and efficient method to dynamically adjust the waiting time during replaying. 
The underlying challenge is to infer GUI rendering states, differentiating between partially rendered and fully rendered GUIs. Inspired by the fact that humans can easily classify these GUIs visually, we propose to identify the GUI rendering states with visual cognitive techniques.

\renewcommand{\arraystretch}{1}
\begin{table}
    \small
    \tabcolsep=0.2cm
	\centering
	\caption{A study of waiting time in replaying.}
	\label{tab:waiting}
	\begin{tabular}{l|c|c|c|c} 
	    \hline
	    \bf{Method} & \bf{1x wait} & \bf{2x wait} & \bf{5x wait} & \bf{10x wait} \\
	      \hline
            \hline
           Reproducibility & 11\% & 33\% & 77\% & 100\% \\
           Time & 23.2s & 49.4s & 112.1s & 217.5s \\
            \hline
	\end{tabular}
\end{table}

\section{\tool Tool}
This paper presents a simple but effective approach for adaptively adjusting the waiting time based on GUI states. 
Drawing inspiration from the previous approach, AdaT~\cite{feng2023efficiency}, which aimed to accelerate automated app testing by categorizing GUI rendering states, we explore its practical implementation in the industrial app WeChat and further expand the approach for recording and replaying.

The overview of our tool is shown in Figure~\ref{fig:overview}.
In the recording phase, we record all input events, including widget attributes, actions, and input data, and parse them into the testing script.
In the replaying phase, we synchronously capture the GUI screenshots and detect their current state.
If the GUI is fully rendered, we schedule the next replaying events; otherwise, we explicitly wait for rendering to complete.
% (i) the \textit{WeChat Data Preparation} phase, which automatically collects a large-scale dataset of partially loaded GUIs and fully loaded GUIs in the WeChat app,
% (ii) the \textit{GUI State Classification} phase, which proposes a CNN-based model to discriminate the WeChat GUI state,
% and (iii) the \textit{Model Implementation} phase that implements our model in record and replay.

\begin{figure}
	\centering
	\includegraphics[width=0.99\linewidth]{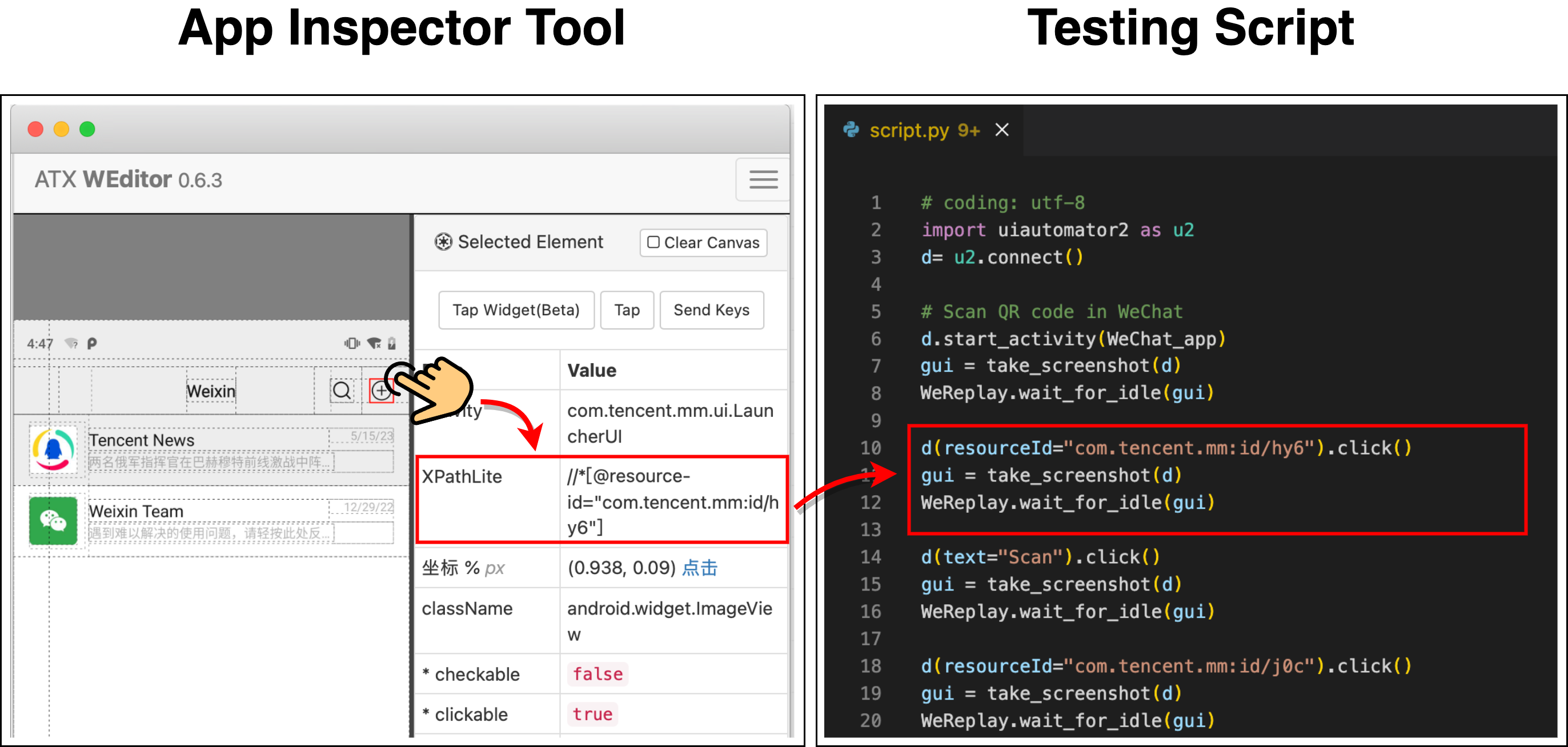}
	\caption{The testing script recorded by app inspector tool.}
	\label{fig:wexintool}
\end{figure}

\subsection{Recording Phase}
There are many available tools for recording testing scenario scripts, either manually~\cite{wang2022characterizing} or automatically~\cite{guo2019sara,zadgaonkar2013robotium}.
To ensure the effectiveness of testing scripts at the industrial level, we manually record the rich variety of events, including the widget attributes, actions, and input text.
Figure~\ref{fig:wexintool} illustrates how a user event is transformed into a script.
In detail, we first use the app inspector tool WEditor~\cite{web:weditor} to inspect the widget information from the interface, as shown in Figure~\ref{fig:wexintool}.
Given the widget information, we then parse it into executable testing scripts using the widely-used automated app testing framework UIAutomator2~\cite{web:uiautomator}.
It is worth noting that other tools can also be employed for this purpose.

\subsection{Replaying Phase}
The foundation of our approach is to utilize a lightweight CNN-based model to classify the WeChat GUI rendering state during replaying. 
To achieve this, we first introduce a novel approach for collecting a large-scale dataset consisting of partially rendered and fully rendered GUIs in the WeChat app. 
Using this data, we propose a CNN-based model to distinguish between WeChat GUI states.

\begin{figure}
	\centering
	\includegraphics[width=0.95\linewidth]{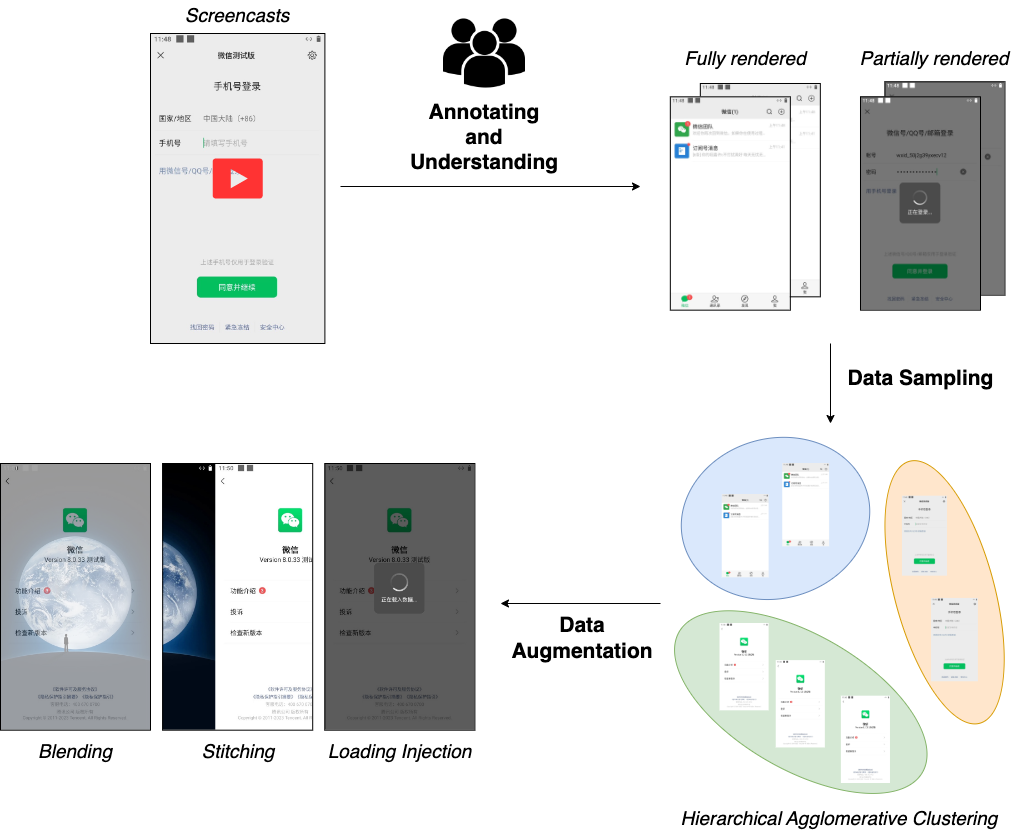}
	\caption{The pipeline of WeChat data preparation.}
	\label{fig:dataset}
\end{figure}

\subsubsection{WeChat Data Preparation}
\label{sec:phase1}
The foundation of training deep learning models is big data.
To begin, we first manually annotate the record-and-replay screencasts from Section~\ref{sec:motivation}, while gaining an understanding of the GUI rendering states in WeChat.
Based on this understanding, we introduce three tailored data augmentation techniques, e.g., stitching, blending, and loading injection, to automatically synthesize more data.
A pipeline for preparing WeChat data is illustrated in Figure~\ref{fig:dataset}.

\textbf{Annotating and understanding.}
We recruited two developers as annotators through WeChat’s internal slack channel.
According to the pre-study background survey, they have previously labeled one UI/UX-related dataset (e.g., GUI element bounding box). 
To ensure accurate annotations, the process started with initial training.
First, we asked them to read the previous GUI rendering study~\cite{feng2023efficiency} and a document~\cite{web:render} that outlines the GUI rendering process. 
Second, we provided an example set of annotated GUIs where the rendering states have been labeled by authors. 
This enforces a deeper understanding of the GUI rendering states. 
Third, we asked them to pass an assessment test, which includes a set of test GUIs.
Finally, we asked them to manually check 23,331 GUIs from 36 record-and-replay screencasts.
During the manual examination process, we identified three types of GUI states following the Card Sorting~\cite{spencer2009card} method:

\begin{figure}
	\centering
	\subfigure[Transiting state]{
		\includegraphics[width = 0.6\linewidth]{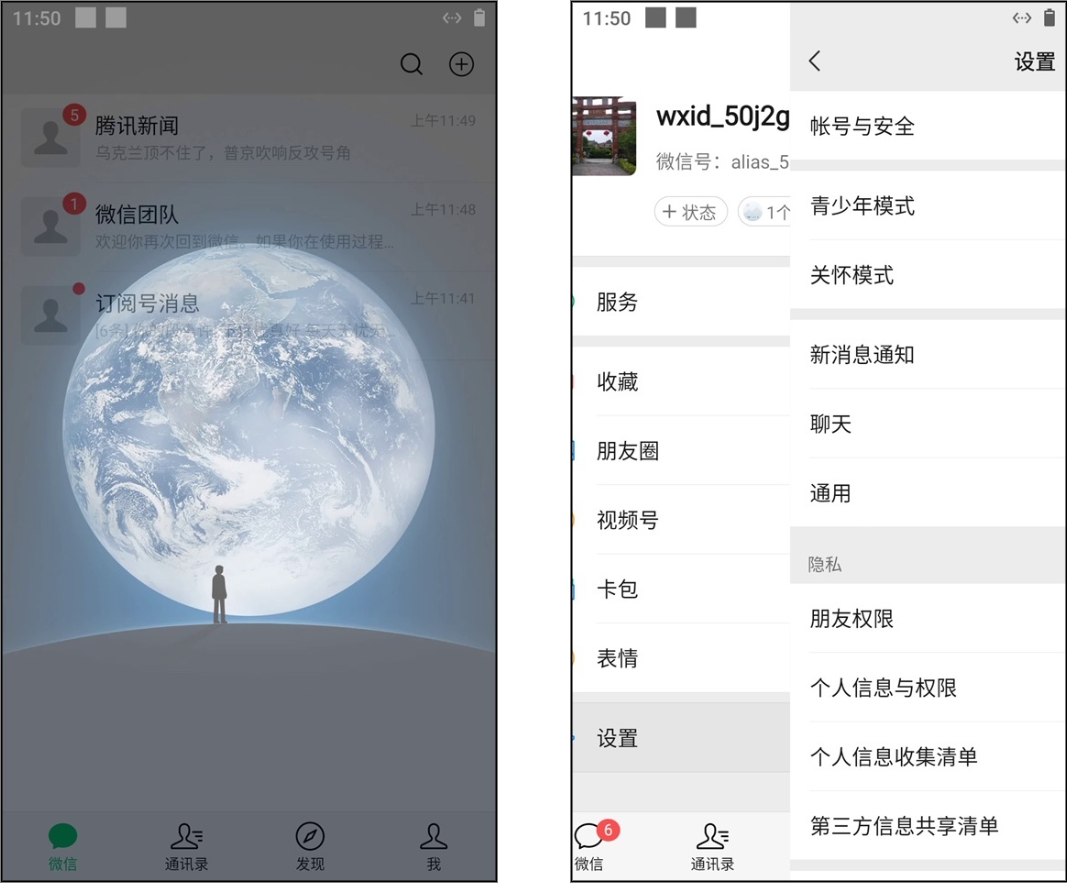}
		\label{fig:partial_1}}
	\hfill
	\subfigure[Loading state]{
		\includegraphics[width = 0.278\linewidth]{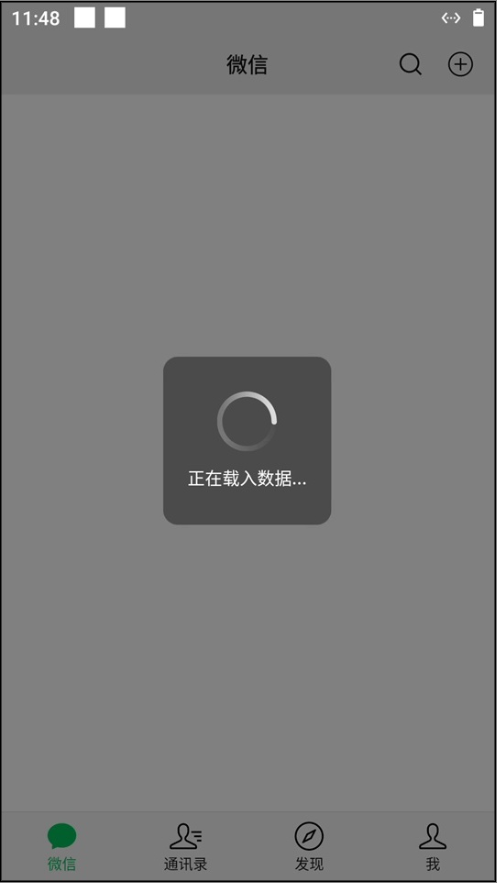}
		\label{fig:partial_2}}	
	\caption{Examples of partially rendered state.}
	\label{fig:partial}
\end{figure}

\textit{a) Transiting State:}
As depicted in Figure~\ref{fig:partial_1}, one state is transiting to the next state.
The transition between states takes time, causing GUIs to overlap or merge during the process. 
This issue often occurs in devices with subpar performance and hardware acceleration defects, resulting in unexpectedly prolonged loading times.
Replaying events in this state may lead to unforeseen errors.

\textit{b) Loading State:}
As illustrated in Figure~\ref{fig:partial_2}, this state displays an explicit loading indicator in the GUI, such as a spinning wheel, linear progressing bar, textual hint, etc.
It explicitly indicates the process or rendering is in progress and is often used for secure data transformations, such as logging into accounts, transferring money, uploading a file, etc.
During the explicitly loading state, the GUI is non-interactive to replay events.

\textit{c) Fully Rendered State:}
A fully rendered state represents a GUI rendered completely with all resources loaded and displayed.
%a complete transition to the GUI with all resources loaded.

\textbf{Data sampling.}
For each GUI state group, we observe that many GUIs are duplicated. 
This is because the rendering changes relatively slowly between consecutive GUI frames from the screencasts. 
% In addition, the same GUIs are triggered across the screencasts, resulting in redundant data in the dataset.
To prevent this bias, we propose a paradigm that uses hierarchical agglomerative clustering (HAC)~\cite{mullner2011modern} to sample the GUI frames from the GUI groups.
% It is chosen because of its novel bottom-up strategy that does not require specifying the number of clusters in advance.

\begin{algorithm}[t]
    % \linespread{0.2}
	\SetAlgoLined
	\SetNoFillComment
	\SetKwInOut{Input}{Input}
	\SetKwInOut{Output}{Output}
	\Input{GUI frames $F = \{f_1, f_2, f_3, ..., f_n\}$; \\
                Agglomerative similarity threshold $\epsilon$; }
	\Output{Sampled frames $S$}
        \tcc{Initialize}
        let $com[0..m, 0..n]$ be new computation table \;
	initialize $com \gets 0$ \;
        \For{$i \gets 1$ to $n$}{
            \For{$j \gets 1$ to $n$}{
                $com[i][j] \gets SSIM(f_i, f_j)$
            }
        }     
        
        $S \gets []$\;
        \tcc{Iteratively merge the new cluster until reaching threshold $\epsilon$}
	\While{$F.length$ > 1}{
	    $(f_i,f_j) \gets argmax(com) $ \;
            $F.remove(\{f_i, f_j\})$ \;
            \uIf{$com[i][j] \geq \epsilon$}{
                $S.append(f_i)$ \;
            }\Else{
                $F.append(f_i \cup f_j)$ \;
            }
            \tcc{Update computation table to new clusters}
            \For{$i \gets 1$ to $F.length$}{
                \For{$j \gets 1$ to $F.length$}{
                    $com[i][j] \gets SSIM(f_i, f_j)$
                }
            }
	}
	return $S$
	\caption{Hierarchical Agglomerative Clustering}
	\label{algorithm:hac}
\end{algorithm}

The detail of HAC is shown in Algorithm~\ref{algorithm:hac}. 
That is, each GUI frame is initially considered as a single-element cluster (Lines 1-7). 
At each step of the algorithm, the two most similar clusters are merged into a new cluster (Line 10).
This procedure is iterated until all the clusters reach the agglomerative similarity threshold $\epsilon$ (Lines 12-16).
 
The main parameter in this algorithm is the metric used to compute the similarity value of GUI frames. 
To achieve this, we adopt the commonly-used perceptual metric, Structural Similarity Index (SSIM)~\cite{wang2004image}, which produces a per-pixel similarity value related to the local difference in average value, variance, and correlation of luminances.
The SSIM score ranges between 0 and 1, with a higher value indicating a strong level of similarity.
Considering the properties of sparseness and distinctness in data sampling, we empirically set the threshold $\epsilon$ to 0.9.
After automated clustering and sampling, we obtain a dataset with 4,485 fully rendered GUIs and 3,159 partially rendered GUIs.

\textbf{Data augmentation.}
Training an effective deep learning model requires a large amount of input data~\cite{lecun2015deep}.
However, gathering and labeling relevant GUIs can be both time-consuming and labor-intensive.
Data augmentation techniques, such as cropping, rotation, color space transformations, etc., are commonly employed to overcome this constraint~\cite{shorten2019survey}.
The closer the synthetic data is to the real one, the better the model's performance. 
Therefore, we develop three novel data augmentation methods based on our aforementioned understanding of the WeChat GUIs.

\textit{a) Stitching:}
As illustrated in Figure~\ref{fig:partial_1}, one particular transiting state occurs when one GUI slides to the next. 
To synthesize this, we employ image stitching, which combines two GUIs with segmented states to create a sliding view. 
Specifically, we randomly crop the GUIs horizontally, as GUI sliding typically occurs in this direction. 
Next, we generate a seamless connection between the two cropped GUIs, resulting in a synthesized GUI.
% as depicted in Figure~\ref{??}.

\textit{b) Blending:}
As illustrated in Figure~\ref{fig:partial_1}, another transiting state occurs when one GUI fades out while the next GUI fades in, causing overlap between the two GUIs. 
To synthesize this, we employ image blending, which linearly combines two GUIs in a gradient transition.
Specifically, we utilize a weighted average method, where the opacity of each GUI is adjusted according to a random transition curve, to ensure a smooth transition between the GUIs.
% A synthesized GUI is depicted in Figure~\ref{??}.

\textit{c) Loading Injection:}
As illustrated in Figure~\ref{fig:partial_2}, the loading state typically displays an explicit loading symbol, such as a spinning wheel.
To synthesize the loading state, we inject the loading symbol into the GUIs.
Additionally, we apply varying intensities of shadow effects to the GUIs to create a more dynamic loading state.
% as depicted in Figure~\ref{??}.

\subsubsection{GUI Rendering State Classification}
\label{sec:phase2}
To identify whether a GUI is fully rendered, allowing the replaying tool to execute the next event, we adopt an implementation of MobileNetV2~\cite{sandler2018mobilenetv2}.
This model distills the best practices in convolutional network design into a simple architecture that offers competitive performance while maintaining low parameters and mathematical operations to reduce computational costs and memory overhead.
% In addition to the simulator, the model can even be deployed on mobile devices for efficient testing.
This advanced network design accelerates image classification, which is the ultimate goal of this work to efficiently discriminate between GUI rendering states.

Specifically, we adopt a more advanced depthwise separable convolution, combining one $3*3$ convolution layer and two $1*1$ convolution layers to capture essential information from images.
We first use a $1*1$ pointwise convolution layer to expand the number of channels in the input feature map.
Then, we use a $3*3$ depthwise convolution layer to filter the input feature map and a $1*1$ convolution layer to reduce the number of channels of the feature map.
In order to improve the performance and stability between layers, we borrow the idea of residual connection in ResNet~\cite{he2016deep} to help with the flow of gradients.
After the convolution layer, we add Batch Normalization (BN)~\cite{ioffe2015batch} to standardize the feature map.
Finally, the activation function, Rectified Linear Unit (ReLU), is added to increase the nonlinear properties of the classifier function and of the overall network without affecting the features.

For detailed implementation, we fine-tune the model from previous work~\cite{feng2023efficiency} using our tailored data to adjust the parameters in the classification layers for the specific industrial app, WeChat.
We adopt the stride of 2 in the depthwise convolution layer to downsample the feature map.
We use ReLU6 defined as $y=min(max(0,x),6)$, for the first two activation layers because of its robustness in low-precision computation~\cite{howard2017mobilenets}.
A linear transformation (also known as Linear Bottleneck Layer)~\cite{sandler2018mobilenetv2} is applied to the last activation layer to prevent ReLU from destroying features.
The momentum in the BN layer is set as 0.1.
To make our training more stable, we adopt Adam as optimizer~\cite{kingma2014adam}, and binary CrossEntropyLoss as the loss function~\cite{murphy2012machine}.
Moreover, to optimize the training model, we apply an adaptive learning scheduler, with an initial rate of 0.01 and decay to half after 10 iterations.
For data preprocessing, we resize the GUI screenshots to $768*448$. 
We implement our model based on the PyTorch framework~\cite{paszke2019pytorch}.
Note that the hyper-parameter settings are determined empirically by a small-scale experiment.

\subsubsection{Model Deployment}
To enable the model to efficiently provide feedback on the GUI rendering state to the replaying tool, synchronization between the GUI and the replaying tool is necessary. 
In detail, we use Android Debug Bridge (adb)~\cite{web:adb} to capture and transmit real-time GUI screenshots. 
Once the screenshot is received, we decode it into a PyTorch tensor~\cite{paszke2019pytorch}. 
This tensor is then fed into our trained GUI state classification model to infer the current GUI's rendering state. 
If the GUI is fully rendered, we proceed to replay the new event; otherwise, we explicitly wait for the next screenshot.
To prevent excessive time consumption due to prolonged resource loading or incorrect model predictions, we set a maximum waiting threshold. 
This waiting threshold is empirically set at 60 seconds based on a small pilot study.

\section{Evaluation}
\label{sec:evaluation}
In this section, we describe the procedure we used to evaluate our approach in terms of its accuracy, effectiveness, and efficiency to accelerate the record-and-replay process.
To achieve our evaluation, we formulate the following three research questions:

\begin{itemize}[leftmargin=0.3cm] 
    \item \textbf{RQ1}: How accurate is our model in classifying WeChat GUI rendering state?
    \item \textbf{RQ2}: How effective and efficient is our tool in replaying WeChat usage scenarios on the same device?
    \item \textbf{RQ3}: How effective and efficient is our tool in replaying WeChat usage scenarios on different devices?
    % \item \textbf{RQ4}: What is the time and space overhead of our approach?
\end{itemize}

For \textbf{RQ1}, we present some general performance of our model in inferring WeChat GUI states and the comparison with state-of-the-art baselines.
For \textbf{RQ2}, we carry out experiments to check if our tool can speed up the automated replaying of the usage scenarios in WeChat on the same device, without sacrificing the effectiveness.
For \textbf{RQ3}, we conduct experiments to evaluate the effectiveness and efficiency of our tool to replay on different devices with diverse screen sizes and operating system versions.

\subsection{RQ1: Performance of WeChat GUI State Classification}
\subsubsection{Experimental Setup.}
To answer RQ1, we first evaluated the ability of our model MobileNetV2 (in Section~\ref{sec:phase2}) to accurately differentiate between fully rendered GUIs and partially rendered GUIs.
To accomplish the evaluation, we followed the procedure to generate the dataset outlined in Section~\ref{sec:phase1}.
We collected 23,331 GUI screencasts and annotated 9,119 partially rendered GUIs and 14,212 fully rendered GUIs.
Due to the presence of duplicates in the dataset, we employed a clustering algorithm to eliminate these redundancies, resulting in 3,159 partially rendered GUIs and 4,485 fully rendered GUIs.
Next, we used tailored data augmentation methods to synthesize 3,012 partially rendered GUIs.
In total, we obtained 6,171 partially rendered GUIs and 4,485 fully rendered GUIs as our experimental dataset.
Regarding our training-testing data split, we split this 10k GUIs in the ratio of 8:1:1 for the training, validation, and testing sets, respectively.
The resulting split has 8k GUIs in the training dataset, 1k GUIs in the validation dataset, and 1k GUIs in the testing dataset.
The model was trained in an NVIDIA GeForce RTX 2080Ti GPU (16G memory) with 20 epochs for about 2 hours.

\subsubsection{Metrics.}
Since we formulated our problem as an image classification task, we adopted three widely-used metrics i.e., precision, recall, and F1-score, to evaluate the accuracy of the model inference.
Precision is the proportion of GUIs that are correctly predicted as fully rendered among all GUIs predicted as fully rendered.

$$ precision = \frac{\#GUIs \ correctly \ predicted \ as \ fully \ rendered}{\#All \ GUIs \ predicted \ as \ fully \ rendered} \\ $$

Recall is the proportion of GUIs that are correctly predicted as fully rendered among all fully rendered GUIs.

$$ recall = \frac{\#GUIs \ correctly \ predicted \ as \ fully \ rendered}{\#All \ fully \ rendered \ GUIs} \\  $$
F1-score (F-score or F-measure) is the harmonic mean of precision and recall, which combine
both of the two metrics above.

$$ F1-score = \frac{2 \times precision \times recall}{precision + recall} $$
For all metrics, a higher value represents better performance.
% Since the ultimate goal is to speed up testing process, we also measured the time for inference.
% For the inference time, a lower time cost represents faster inference of the GUI rendering state.

% The higher the accuracy score, the better the model can discriminate the GUI rendering state.
% To prevent additional computation cost from the invocation of initializing GPU~\cite{web:warmup}, we warmed up the GPU by randomly running over 1000 samples.

\renewcommand{\arraystretch}{1.05}
\begin{table}
    \small
    \tabcolsep=0.35cm
    \centering
	\caption{Performance comparison with baselines}
	\label{tab:rq1}
	\begin{tabular}{l|c|c|c}
	\hline
	\bf{Methods} & \bf{Precision} & \bf{Recall} & \bf{F1-score}\\ 
	\hline
        SIFT+SVM & 0.706 & 0.751 & 0.727\\
    SIFT+KNN & 0.592 & 0.639 & 0.614\\
	SIFT+RF & 0.678 & 0.670 & 0.673\\
	\hline
	SURF+SVM & 0.634 & 0.625 & 0.629 \\
    SURF+KNN & 0.577 & 0.601 & 0.588\\
	SURF+RF & 0.606 & 0.659 & 0.631\\
	\hline
    ORB+SVM & 0.659 & 0.699 & 0.678\\
    ORB+KNN & 0.596 & 0.616 & 0.605\\
	ORB+RF & 0.636 & 0.667 & 0.651\\
	\hline
    % CNN & 0.813 & 0.791 & 0.801\\
    AdaT & 0.859 & 0.852 & 0.855 \\
	% \hline
        \tool w/o aug & 0.893 & 0.906 & 0.899 \\
	\tool & \bf{0.921} & \bf{0.933} & \bf{0.927} \\
	\hline
	\end{tabular}
\end{table}

\subsubsection{Baselines.}
We set up 9 state-of-the-art baseline methods, that are widely used in image classification tasks as the baselines to compare with our model.
They first extract visual features from the GUI screenshots and then employ a machine learner for the classification.
In detail, we adopted three types of feature extraction methods used in machine learning, e.g., Scale-invariant feature transform (SIFT)~\cite{lowe2004distinctive}, Speed up robot features (SURF)~\cite{bay2006surf}, and Oriented fast and rotated brief (ORB)~\cite{rublee2011orb}.
With these features, we applied three commonly-used machine learning classifiers, e.g., Support Vector Machine (SVM)~\cite{kotsiantis2007supervised}, K-Nearest Neighbor (KNN)~\cite{keller1985fuzzy}, and Random Forests (RF)~\cite{breiman2001random}, for classifying the GUI rendering state.
The combination of three types of image features and three classification learning algorithms generated a total of 9 baselines.

In addition, we also added 2 ablation studies as our baselines to demonstrate the advantage of our model.
% In addition, we also add 2 derivatives of our approach to demonstrate the impact of each component. 
First, we experimented with the off-the-shelf pre-trained model AdaT~\cite{feng2023efficiency} to see its general performance in WeChat GUIs.
It uses a convolutional neural network MobileNetV2 to extract the visual features and trains on 79k GUIs from the 9.7k Google app to classify the fully rendered and partially rendered states.
Second, we investigated the contribution of our data augmentation methods in Section~\ref{sec:phase1}, namely \tool w/o aug, to see the performance of our model trained without 3,012 (40\%) additional data.

\subsubsection{Result.}
\label{sec:rq1_result}
Table~\ref{tab:rq1} illustrates the performance of our approach in classifying the fully rendered GUIs and partially rendered GUIs in the industrial app WeChat.
Our approach significantly outperforms other baselines, achieving a 28.7\%, 30.8\%, and 29.8\% increase in recall, precision, and F1-score, respectively, compared to the best machine learning baseline (SIFT+SVM).
We observe that deep learning-based methods perform much better than machine learning methods, primarily because machine learning lacks feature introspection, which is crucial as GUI rendering state features vary.

Compared to the pre-trained deep learning baseline AdaT, our model further improves recall, precision, and F1-score by 8.1\%, 6.2\%, and 7.2\%, respectively. 
This suggests that fine-tuning a pre-trained model enables it to better recognize specific features, such as the distinct GUI characteristic in the industrial app WeChat. 
Consequently, this enhances the model's ability to classify GUI states more accurately. 
Additionally, we observe that applying tailored data augmentation methods further improves our model's performance, increasing recall, precision, and F1-score by 2.7\%, 2.8\%, and 2.8\%, respectively. 
This indicates that with more training data, the model's GUI rendering classification capability improves. 
In the future, we plan to collect more diversified GUIs from the WeChat app to enhance the model's performance.

\begin{figure}
	\centering
	\subfigure[]{
		\includegraphics[width = 0.225\linewidth]{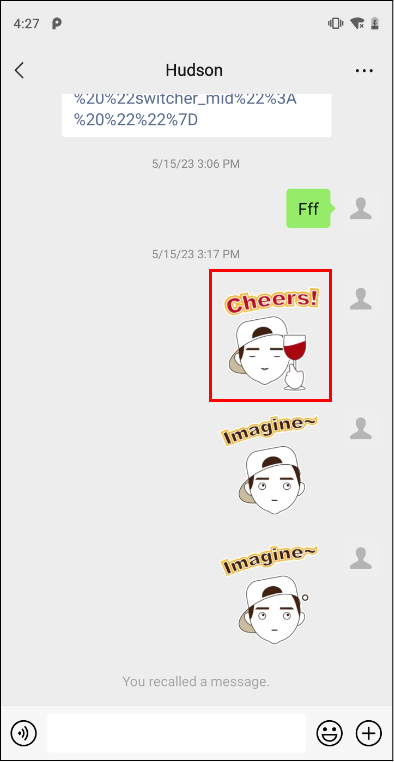}
		\label{fig:eval1_1}}
	\hfill
	\subfigure[]{
		\includegraphics[width = 0.225\linewidth]{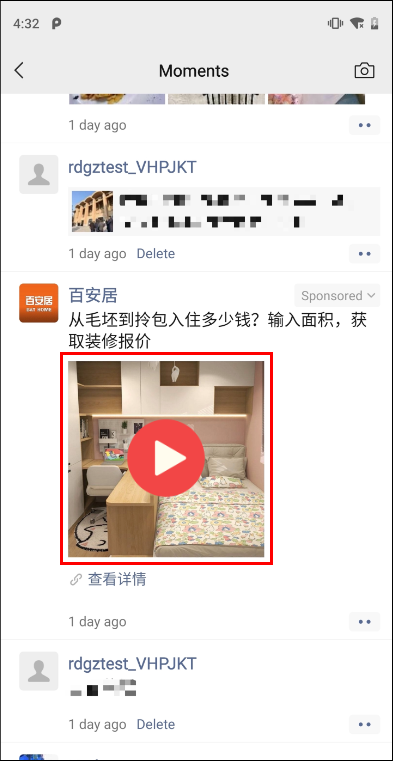}
		\label{fig:eval1_2}}
        \hfill
	\subfigure[]{
		\includegraphics[width = 0.225\linewidth]{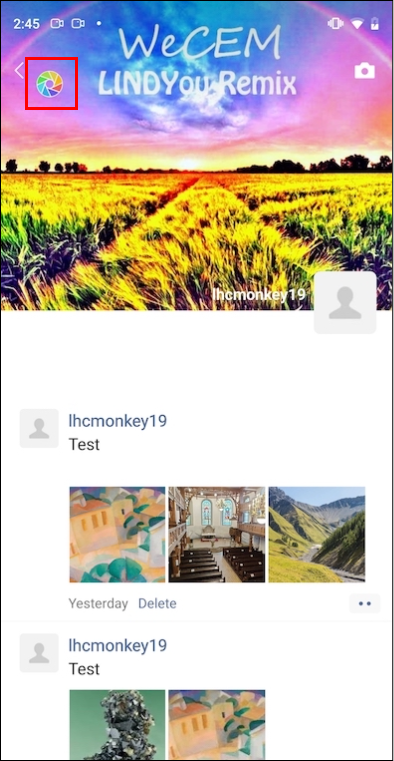}
		\label{fig:eval1_3}}
        \hfill
	\subfigure[]{
		\includegraphics[width = 0.225\linewidth]{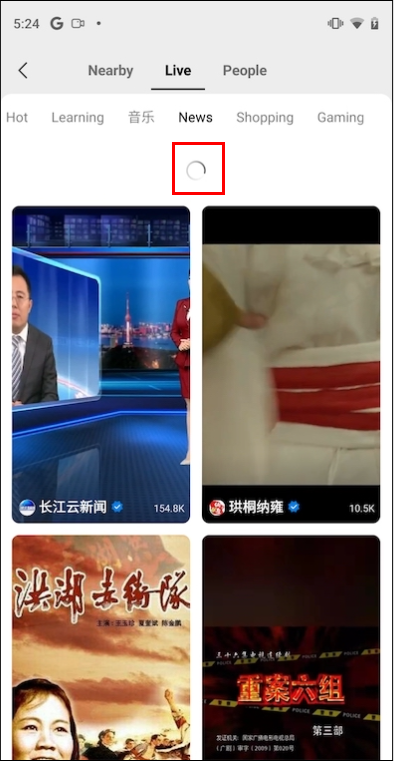}
		\label{fig:eval1_4}}
	\caption{Examples of bad cases in GUI state prediction.}
	\label{fig:partial}
\end{figure}

Albeit the good performance of our model, we still make incorrect predictions for some GUI screenshots. 
We manually examined these inaccuracies and identified two common causes.
First, within the industrial app WeChat, GUIs may contain dynamic assets such as videos and gifs, as seen in the gif message in Figure~\ref{fig:eval1_1} and the advertisement in Figure~\ref{fig:eval1_2}. 
In these cases, although the GUIs are fully rendered, they may be misclassified as partially rendered GUIs due to animation loading.
Second, some representative features of loading in WeChat are too small and inconspicuous to be recognized, even by human eyes, for example, the tiny circular progress bar embedded in the image in Figure~\ref{fig:eval1_3} and embedded in the background in Figure~\ref{fig:eval1_4}.

\subsection{RQ2: Performance of Recording and Replaying on Same Device}
\subsubsection{Experimental Setup.}
\label{sec:rq2_p1}
To answer RQ2, we evaluated the ability of our tool to effectively and efficiently record and replay usage scenarios on the same device.
Throughout our three months study, we had access to WeChat developers and periodically asked them for the scenarios on their hands-on experiences using the app.
In total, we gathered 23 fundamental usage scenarios in the WeChat app as our experimental dataset.
On average, each scenario contained 5.3 events, covering various WeChat functionalities such as messaging, games, shopping, etc.
Note that we did not use the usage scenarios collected in Section~\ref{sec:motivation} for this evaluation, as they were utilized for model training and could potentially result in a data leakage problem~\cite{kaufman2012leakage}.
% To avoid potentail bias, we further audited the experimental dataset to guartee

We recorded and replayed the usage scenarios on Xiaomi Mix2S running Android version 9.0.
The device was allocated with 8 dedicated CPU cores, 6GiB of RAM, and discrete graphics cards for minimal mutual influences caused by disk I/O bottlenecks and CPU-intensive graphical rendering.

\subsubsection{Metrics.}
To measure the performance of our approach, we employed reproducibility as the evaluation metric, i.e., whether the method can successfully replay the usage scenarios. 
The higher the reproducibility score, the better the approach can replicate the events to reproduce the scenarios.
Since the ultimate goal is to speed up the replaying process, we also measured the time for replaying.
For the replaying time, a lower time cost represents faster replaying of the recorded scenarios.

\subsubsection{Baselines.}
We set up two methods as our baseline to compare with our approach.
First, we adopted the state-of-the-art method \textit{SARA}~\cite{guo2019sara}.
Specifically, SARA employs a heuristic technique to record the waiting time elapsed between events.
Second, we followed the practical solution in WeChat, as discussed in Section~\ref{sec:motivation}, extending the waiting time by 10 times to create an industrial baseline, referred to as \textit{10x wait}.
Note that we did not evaluate other publicly available tools like Espresso~\cite{web:espresso} and Culebra~\cite{web:culebra}, as they are not sensitive to waiting time, e.g., setting arbitrary inter-event time, which prevents effective evaluation on industrial apps.

\subsubsection{Result.}
\label{sec:rq2_result}
Table~\ref{tab:rq2_performance} shows detailed results of the time and reproducibility for each usage scenario.
On average, our tool \tool takes 18.45 seconds to reproduce all the scenarios.
We observe that, on one hand, SARA, with a short fixed waiting time, can only replay 39.1\% of the scenarios.
The failure cases are due to dynamic Internet loading, i.e., resources taking longer to load than the recorded waiting time.
On the other hand, the industrial solution of extending the waiting time by 10 times significantly slows down the replaying process, taking an average of 152.99 seconds per scenario.
In contrast, our tool can replay all scenarios (100\%) in less time (18.45 seconds), achieving 60.9\% more successful reproductions compared to SARA, while saving 88\% more time than 10x wait with the same reproduction rate.
As a result, \tool can expedite the replaying process without sacrificing reproducibility, saving a significant amount of time in long-term industrial testing involving hundreds or thousands of events.

% Even for some scenarios that takes longer time cannot successfully replay is due to the key event waiting time is not sufficient.

\renewcommand{\arraystretch}{1.05}
\begin{table}
    \small
    \tabcolsep=0.04cm
	\centering
	\caption{Performance comparison of replaying on the same device. ``R'' denotes scenario reproducibility. ``T'' denotes the time to replay in seconds. }
	\label{tab:rq2_performance}
	\begin{tabular}{l|r||c|c|c|c|c|c} 
	    \hline
	    \multirow{2}{*}{\bf{Scenario}} & \multirow{2}{*}{\bf{Events}} & \multicolumn{2}{c|}{\bf{SARA}} & \multicolumn{2}{c|}{\bf{10x wait}} & \multicolumn{2}{c}{\bf{\tool}} \\
	    \cline{3-8}
	     & & R & T & R & T & R & T \\
	     \hline
           \rowcolor{lightgray} Open Moments & 3 & \cmark & 15.62 & \cmark & 149.13 & \cmark & 12.39 \\
            Open GameCenter & 4 & \cmark & 23.16 & \cmark & 229.61 & \cmark & 22.21 \\
            \rowcolor{lightgray} Open Channels & 4 & \cmark & 22.01 & \cmark & 224.33 & \cmark & 20.67 \\
            Open e-Wallet & 4 & \xmark & 10.20 & \cmark & 96.45 & \cmark & 18.96 \\
            \rowcolor{lightgray} Open Pay QRCode & 4 & \xmark & 10.92 & \cmark & 101.17 & \cmark & 13.32 \\
            Open Collections & 4 & \cmark & 10.13 & \cmark & 98.36 & \cmark & 9.36 \\
            \rowcolor{lightgray} Search Info & 6 & \xmark & 16.24 & \cmark & 161.74 & \cmark & 36.07 \\
            Search Game & 6 & \xmark & 56.96 & \cmark & 553.21 & \cmark & 40.12 \\
            \rowcolor{lightgray} Search New Friend & 6 & \xmark & 12.81 & \cmark & 113.25 & \cmark & 20.37 \\
            Delete Friend & 5 & \cmark & 21.89 & \cmark & 203.93 & \cmark & 14.20 \\
            \rowcolor{lightgray} Enter Shopping & 4 & \cmark & 16.35 & \cmark & 159.01 & \cmark & 12.32 \\
            Recommend Friend & 6 & \xmark & 9.51 & \cmark & 90.43 & \cmark & 12.07 \\
            \rowcolor{lightgray} Change Username & 8 & \xmark & 13.92 & \cmark & 133.86 & \cmark & 15.57 \\
            Subscribe Stickers & 5 & \xmark & 9.11 & \cmark & 89.70 & \cmark & 23.94 \\
            \rowcolor{lightgray} Search Stickers & 8 & \xmark & 15.72 & \cmark & 155.71 & \cmark & 30.88 \\
            Change Profile Photo & 6 & \cmark & 23.82 & \cmark & 219.95 & \cmark & 24.63 \\
            \rowcolor{lightgray} Video Call & 5 & \xmark & 6.52 & \cmark & 64.49 & \cmark & 10.07 \\
            Audio Call & 5 & \xmark & 6.21 & \cmark & 56.62 & \cmark & 9.22 \\
            \rowcolor{lightgray} Unfollow Official Account & 7 & \xmark & 12.42 & \cmark & 117.37 & \cmark & 23.63 \\
            Group Chat & 6 & \xmark & 10.26 & \cmark & 99.67 & \cmark & 14.19 \\
            \rowcolor{lightgray} Send Message & 5 & \cmark & 15.03 & \cmark & 143.44 & \cmark & 12.21 \\
            Send Stickers & 5 & \cmark & 11.79 & \cmark & 112.88 & \cmark & 11.84 \\
            \rowcolor{lightgray} Send Picture & 7 & \xmark & 15.07 & \cmark & 144.46 & \cmark & 16.19 \\
            \hline
            \hline
            Average & 5.35 & 39.1\% & 15.89 & 100\% & 152.99 & \bf{100\%} & \bf{18.45} \\
            \hline
	\end{tabular}
\end{table}
\subsection{RQ3: Performance of Recording and Replaying on Different Devices}
\subsubsection{Experimental Setup.}
To answer RQ3, we evaluated the ability of our tool to effectively and efficiently record and replay events on different devices.
In detail, we used the Xiaomi Mix2S to record the 23 fundamental usage scenarios in the WeChat app outlined in Section~\ref{sec:rq2_p1}.
To replay these usage scenarios, we employed three different devices, including Huawei Nova2S, Vivo Y3, and Google Pixel4a.
Details of the device information can be seen in Table~\ref{tab:device}, covering diverse screen resolutions, sizes, operating systems, and processors.

\subsubsection{Metrics.}
To measure the performance of our tool, we employed two evaluation metrics, i.e., reproducibility (replay success) and time (replay time).

\subsubsection{Baselines.}
We adopted the state-of-the-art \textit{SARA}~\cite{guo2019sara} and industrial solution \textit{10x wait} in Section~\ref{sec:motivation} as the baselines to compare with our tool.

\subsubsection{Result.}
\label{sec:rq3_result}
Table~\ref{tab:rq3_performance} presents the detailed results of the replaying performance across the three devices.
Our tool \tool, successfully replays all (100\%) of the scenarios in a shorter time, with a median of 19.55 seconds.
In contrast, the baseline SARA only replays 21.7\%, 34.8\%, and 47.8\% of the scenarios for Huawei Nova2S, Vivo Y3, and Google Pixel4a, respectively.
The failure scenarios are due to the delayed rendering process for the devices, indicating that more time is needed for rendering.
The industrial solution can also replay all of the scenarios, but it takes much longer, with an average of 157.41 seconds, making it 10 times slower than our tool.
This is because it indiscriminately extends the waiting time between events by 10 times, while most of this time is spent idly waiting.
In contrast, our tool \tool utilizes a deep learning model to identify the GUI rendering state, dynamically scheduling events if the GUI is fully rendered, or explicitly waiting otherwise, achieving efficiency without compromising replay capability.

% Finally, as discussed in Sec. 5.3, one limitation that affects the
% ability of V2S to faithfully replay swipes is the video frame-rate.
% During the evaluation, our devices were limited to 30fps, which
% made it difficult to completely resolve a small subset of gesture
% actions that were performed very quickly. However, this limitation
% could be addressed by improved Android hardware or software capable of recording video at or above 60fps, which, in our experience,
% should be enough to resolve nearly all rapid user gestures.

\renewcommand{\arraystretch}{1.05}
\begin{table*}
    \small
    \tabcolsep=0.085cm
	\centering
	\caption{Performance comparison of replaying on the different devices. ``R'' denotes scenario reproducibility. ``T'' denotes the time to replay in seconds. }
	\label{tab:rq3_performance}
	\begin{tabular}{l||c|c|c|c|c|c|c|c|c|c|c|c|c|c|c|c|c|c} 
	    \hline
	    \multirow{3}{*}{\bf{Scenario}} & \multicolumn{6}{c|}{\bf{Huawei Nova2S}} & \multicolumn{6}{c|}{\bf{Vivo Y3}} & \multicolumn{6}{c}{\bf{Google Pixel4a}} \\
	    \cline{2-19}
	     & \multicolumn{2}{c|}{SARA} & \multicolumn{2}{c|}{10x wait} & \multicolumn{2}{c|}{\tool} & \multicolumn{2}{c|}{SARA} & \multicolumn{2}{c|}{10x wait} & \multicolumn{2}{c|}{\tool} & \multicolumn{2}{c|}{SARA} & \multicolumn{2}{c|}{10x wait} & \multicolumn{2}{c}{\tool} \\
            \cline{2-19}
            & R & T & R & T & R & T & R & T & R & T & R & T & R & T & R & T & R & T \\
	     \hline
           \rowcolor{lightgray} Open Moments & \cmark & 18.77 & \cmark & 150.14 & \cmark & 18.99 & \cmark & 16.81 & \cmark & 147.63 & \cmark & 13.79 & \cmark & 16.31 & \cmark & 145.56 & \cmark & 12.73 \\
            Open GameCenter & \xmark & 27.96 & \cmark & 241.14 & \cmark & 25.58 & \xmark & 25.72 & \cmark & 237.77 & \cmark & 24.91 & \cmark & 16.12 & \cmark & 235.52 & \cmark & 22.37 \\
            \rowcolor{lightgray} Open Channels & \cmark & 28.64 & \cmark & 243.01 & \cmark & 24.64 & \cmark & 24.73 & \cmark & 231.12 & \cmark & 20.97 & \cmark & 24.97 & \cmark & 232.44 & \cmark & 18.65 \\
            Open e-Wallet & \xmark & 13.94 & \cmark & 104.91 & \cmark & 16.36 & \xmark & 12.65 & \cmark & 103.49 & \cmark & 16.33 & \xmark & 11.48 & \cmark & 101.76 & \cmark & 13.95 \\
            \rowcolor{lightgray} Open Pay QRCode & \xmark & 13.01 & \cmark & 107.99 & \cmark & 21.23 & \xmark & 11.97 & \cmark & 107.89 & \cmark & 18.08 & \xmark & 11.50 & \cmark & 104.97 & \cmark & 16.17 \\
            Open Collections & \xmark & 12.75 & \cmark & 108.76 & \cmark & 15.52 & \cmark & 11.03 & \cmark & 104.99 & \cmark & 10.19 & \cmark & 10.93 & \cmark & 103.65 & \cmark & 8.89 \\
            \rowcolor{lightgray} Search Info & \xmark & 20.46 & \cmark & 169.02 & \cmark & 40.96 & \xmark & 18.15 & \cmark & 168.63 & \cmark & 34.66 & \xmark & 17.31 & \cmark & 163.47 & \cmark & 20.86 \\
            Search Game & \xmark & 67.78 & \cmark & 564.98 & \cmark & 51.03 & \xmark & 61.51 & \cmark & 559.98 & \cmark & 44.69 & \xmark & 18.79 & \cmark & 555.72 & \cmark & 29.90 \\
            \rowcolor{lightgray} Search New Friend & \xmark & 16.77 & \cmark & 122.56 & \cmark & 20.24 & \xmark & 13.22 & \cmark & 118.75 & \cmark & 18.09 & \cmark & 14.79 & \cmark & 114.02 & \cmark & 13.98 \\
            Delete Friend & \cmark & 23.75 & \cmark & 215.97 & \cmark & 19.52 & \cmark & 22.19 & \cmark & 210.60 & \cmark & 16.38 & \cmark & 20.01 & \cmark & 203.19 & \cmark & 15.11 \\
            \rowcolor{lightgray} Enter Shopping & \cmark & 19.33 & \cmark & 164.98 & \cmark & 14.07 & \cmark & 17.59 & \cmark & 164.34 & \cmark & 11.77 & \cmark & 18.89 & \cmark & 162.06 & \cmark & 11.15 \\
            Recommend Friend & \xmark & 12.65 & \cmark & 95.09 & \cmark & 16.94 & \xmark & 11.21 & \cmark & 93.68 & \cmark & 12.96 & \xmark & 10.46 & \cmark & 92.18 & \cmark & 10.78 \\
            \rowcolor{lightgray} Change Username & \xmark & 15.89 & \cmark & 141.40 & \cmark & 17.50 & \cmark & 15.92 & \cmark & 136.04 & \cmark & 15.12 & \cmark & 14.80 & \cmark & 135.88 & \cmark & 13.51 \\
            Subscribe Stickers & \xmark & 14.75 & \cmark & 91.57 & \cmark & 31.44 & \xmark & 10.14 & \cmark & 89.94 & \cmark & 26.01 & \xmark & 9.14 & \cmark & 89.07 & \cmark & 26.91 \\
            \rowcolor{lightgray} Search Stickers & \xmark & 17.88 & \cmark & 163.18 & \cmark & 35.78 & \xmark & 17.04 & \cmark & 150.61 & \cmark & 32.06 & \xmark & 16.50 & \cmark & 148.82 & \cmark & 30.85 \\
            Change Profile Photo & \xmark & 29.76 & \cmark & 228.95 & \cmark & 27.90 & \xmark & 26.55 & \cmark & 223.74 & \cmark & 24.49 & \xmark & 25.05 & \cmark & 220.99 & \cmark & 21.66 \\
            \rowcolor{lightgray} Video Call & \xmark & 8.93 & \cmark & 71.93 & \cmark & 20.74 & \xmark & 7.22 & \cmark & 68.62 & \cmark & 10.98 & \xmark & 6.52 & \cmark & 67.02 & \cmark & 7.60 \\
            Audio Call & \xmark & 9.11 & \cmark & 62.07 & \cmark & 16.77 & \xmark & 7.23 & \cmark & 60.49 & \cmark & 9.69 & \cmark & 6.93 & \cmark & 55.48 & \cmark & 7.72 \\
            \rowcolor{lightgray} Unfollow Official Account & \xmark & 17.36 & \cmark & 126.75 & \cmark & 25.04 & \xmark & 14.12 & \cmark & 120.88 & \cmark & 23.04 & \xmark & 13.57 & \cmark & 112.27 & \cmark & 19.20 \\
            Group Chat & \xmark & 12.88 & \cmark & 102.97 & \cmark & 17.86 & \xmark & 11.25 & \cmark & 100.47 & \cmark & 17.53 & \xmark & 11.67 & \cmark & 99.68 & \cmark & 15.96 \\
            \rowcolor{lightgray} Send Message & \cmark & 19.02 & \cmark & 144.98 & \cmark & 17.72 & \cmark & 15.31 & \cmark & 144.44 & \cmark & 12.64 & \cmark & 14.54 & \cmark & 142.52 & \cmark & 12.64 \\
            Send Stickers & \xmark & 14.41 & \cmark & 120.74 & \cmark & 16.83 & \cmark & 13.01 & \cmark & 119.78 & \cmark & 11.81 & \cmark & 12.59 & \cmark & 117.04 & \cmark & 11.32 \\
            \rowcolor{lightgray} Send Picture & \xmark & 17.79 & \cmark & 151.35 & \cmark & 17.93 & \xmark & 15.19 & \cmark & 148.74 & \cmark & 15.89 & \xmark & 15.07 & \cmark & 142.28 & \cmark & 15.63 \\
            \hline
            % \rowcolor{darkgray} 
            \hline
            Average & 21.7\% & 19.72 & 100\% & 160.62 & \textbf{100\%} & \textbf{23.07} & 34.8\% & 17.38 & 100\% & 157.46 & \textbf{100\%} & \textbf{19.17} & 47.8\% & 14.69 & 100\% & 154.16 & \textbf{100\%} & \textbf{16.41} \\
            \hline
	\end{tabular}
\end{table*}

% \subsection{RQ4: Industrial Impact}
% \sidong{See if there is some real impact in industrial, e.g., save money, facilitate testing}
\section{Discussion}
% We had discussed the limitations of our approach at the end of
% each subsection of the evaluation in Section~\ref{sec:evaluation}, such as misclassification due to dynamic assets (Section~\ref{sec:rq1_result}), ?? (Section ??), etc. 
We had discussed the limitations of our tool in misclassifying the WeChat GUI rendering state due to dynamic assets (Section~\ref{sec:rq1_result}).
In this section, we discuss the industrial implication and threats to validity of our tool.

\subsection{Industrial Implication}
Although many state-of-the-art automated recording tools exist~\cite{zadgaonkar2013robotium,guo2019sara,amalfitano2014mobiguitar}, we chose manual recording for usage scenarios in the industrial app WeChat due to two practical lessons learned.
First, automated recording of widgets in hybrid apps like WeChat (apps built with a combination of native and web technologies) presents steering challenges, as some widgets are rendered by WebView as HTMLElements, which cannot be found in the view hierarchy.
Second, games in WeChat's mini-programs~\cite{liu2020industry} are often developed with game engines (e.g., Unity~\cite{web:unity}), making it impossible to automatically record the widget.
To ensure the effectiveness of recording usage scenarios in the WeChat app, we generate them manually by inspecting the widget in the inspector tool WEditor.
In the future, we plan to conduct a comprehensive empirical study of the cases where the widgets cannot be automatically recorded and will attempt to develop an engineering effort to address this issue.

Another industrial implication involves supporting record and replay for different platforms, such as iOS.
The results in Section~\ref{sec:rq2_result} and Section~\ref{sec:rq3_result} demonstrate strong performance in recording and replaying on Android devices.
Although we focus on the Android platform in this study for brevity, our approach could be extended to other platforms.
As the GUI rendering process in iOS exhibits minimal differences compared to Android, our approach could be adapted to it with a reasonable amount of engineering effort.

\subsection{Threats to Validity}
Threats to internal validity may arise from the manual labeling of the training and testing dataset for the GUI rendering classification model.
To mitigate any potential subjectivity or errors, we provided the annotators with a training session and a qualifying test before labeling.
We instructed them to independently annotate without any discussion, and we reached a consensus on the finalized dataset. 
Although there may still be some noise in the data, training a deep learning-based model with a sufficient amount of high-quality data can tolerate a small amount of noise~\cite{sukhbaatar2014training,lecun2015deep}.

In our experiments evaluating our tool, threats to external validity may arise from the representativeness of industrial usage scenarios depicted in our experimental set.
To mitigate this threat, we conducted experiments on 23 real-world usage scenarios provided by WeChat developers.
While performing additional experiments with more scenarios would be ideal, our experimental set of scenarios represents a reasonably fundamental set of tests with different functionalities, illustrating the relative performance of our tool.
Another potential confounding factor concerns the representativeness of the record-and-replay devices used in the evaluation.
To mitigate this threat, we employed four devices with different resolutions, sizes, operating systems, and processors, as outlined in Section~\ref{sec:motivation}.
These devices are practically used as testing devices in WeChat and widely studied in previous studies~\cite{yu2021layout,romano2021empirical}.
% These devices are widely-studied in the previous studies~\cite{} and be practical used as testing devices in WeChat.
One more potential threat concerns the generalizability of our approach to other industrial apps.
In this study, we focus on the WeChat app, using WeChat GUIs to train the GUI rendering classification model to help speed up the record and replay process in practice.
Our approach relies only on GUI screenshots, which should be easily adapted to other industrial apps with customized datasets.
\section{Related Work}
The main quality of our study is the utilization of the GUI rendering state to accelerate the record-and-replay process in the industrial app WeChat.
Therefore, we review the related work in two main areas: 1) record and replay for industrial apps, and 2) efficiency support for testing.

\subsection{Record and Replay for Industrial Apps}
The primary goal of record-and-replay tools is to record an app's execution to facilitate automatic replay.
RERAN~\cite{gomez2013reran} is one of the earliest record-and-replay tools that use the Linux kernel.
Specifically, it captures low-level events with the ADB command getevent by reading logs in \textit{/dev/input/event*} files and uses the command sendevent to replay events. 
However, the low-level events recorded are tightly coupled to the hardware, making it difficult to reconstruct high-level gestures for replay, such as zoom, pinch-in, etc.
MobiPlay~\cite{qin2016mobiplay} introduces a client-server architecture, involving a client app running on a mobile device and a target app running on the server to record events.
It requires a custom OS to record and replay, which may lead to incompatibilities between the OS and the device, thereby violating industry constraints.
Monkeyrunner~\cite{web:monkeyrunner}, a desktop tool developed by Google Inc., provides an API for writing Python programs to record and replay keystroke coordinates on Android devices.
Subsequent tools like Mosaic~\cite{halpern2015mosaic}, HiroMacro~\cite{web:hiroMacro}, VALERA~\cite{hu2015versatile}, and RepetiTouch~\cite{web:repetiTouch}  capture events in a similar manner.
Nevertheless, the underlying recording and replaying based on pixel coordinates are often prone to failure due to minor GUI changes across diverse devices.
% Other pixel-based approaches include VALERA [26] (a customized
% system) and Mobiplay [44] (record and replay on a remote-desktop).

Consequently, several researchers~\cite{feng2023prompting,feng2023read} have developed widget-sensitive record-and-replay tools.
For example, Robotium~\cite{zadgaonkar2013robotium} is derived from the Selenium web browser automation tool and records events only if GUIs are controlled by the app's main process.
Whereas, many industrial apps may use different processes to avoid compatibility issues on various platforms.
One example of how WeChat uses other processes to create GUI widgets is through its mini-programs, which are embedded frameworks within the WeChat app.
Ranorex~\cite{web:ranorex} is a commercial test automation tool that supports recording events based on widgets through app instrumentation.
However, instrumentation requires sophisticated accessibility or GUI automation APIs~\cite{feng2022gifdroid,feng2022gifdroid2} and continuous updates in sync with the app and different operating systems~\cite{li2018cid,wang2013detecting}, making it incompatible with industrial apps.
The official Android Studio IDE introduces Espresso~\cite{web:espresso}, which leverages source code analysis to record events based on widgets by attaching a debugger to the app, but it still requires developers to manually set the time delay of events, which can be troublesome and error-prone.

Guo et al.~\cite{guo2019sara} introduce a practical record-and-replay tool SARA, that satisfies the industry requirements for widget-sensitive and time-sensitive recording and replaying.
Specifically, it proposes a self-replay mechanism to record user event information while capturing timestamps to infer the waiting time between replay events.
However, a fixed waiting time may not accurately replay events according to our analysis (less than 55\% for the same device and 37\% for different devices) in Section~\ref{sec:motivation}.
First, the waiting time can be indeterminate, depending on the internet, which is frequently used in industrial apps like WeChat. 
Second, for cross-device replaying, the waiting time can be significantly dependent on the performance of the devices, with lower-performing devices typically requiring more waiting time.
In contrast, our tool does not record a fixed waiting time but leverages a novel deep learning-based model to infer GUI rendering state, dynamically adjusting the waiting time to schedule events on fully rendered GUIs.
The empirical evaluations of record and replay in the WeChat app confirm the practicality of our tool.

% Reran, Mosiac and  -> linux kernel
% MobiPlay -> custom OS
% appetizer -> coordinate-sensitive
% Robotium, Ranorex -> app instrumentation
% Culebra -> impaticality
% Espresso -> need source code, not time sensitive
% VALERA
% HiroMacro, RepetiTouch -> need root access
% SARA -> not very time

% There are also many related works on record and replay using supplementary information.
% Some researchers~\cite{zhao2019recdroid, fazzini2018automatically,zhao2019automatically,song2020bee,thummalapenta2012automating} leverage the natural language processing methods to replay the bug scenarios from the textual information in bug reports, while some researchers~\cite{??} focus on replaying  from the visual information using image processing and deep learning methods.
% We assume these approaches are complementary to our approach, that uses mulit-modal information to potentially improve the reproducibility of LLMs' understanding. 

\subsection{Efficiency Support for Testing}
Many works have attempted to improve infrastructure support for efficient mobile testing.
Hu et al.~\cite{hu2014efficiently} propose AppDoctor, which instruments target apps using event handler invocations to quickly identify potential sequences of error-triggering GUIs.
Song et al.~\cite{song2017ehbdroid} enhance the efficiency of AppDoctor by leveraging direct invocations. 
Wang et al.~\cite{wang2021infrastructure} propose an Android tool, Toller, that injects into the testing device to efficiently access GUI layout and execute events.
In contrast to these infrastructure support methods, our goal is to accelerate record and replay by using adaptive waiting times, i.e., scheduling testing events for efficiency improvement.

Adaptive waiting time is a common practice for efficient record and replay on the web.
Selenium~\cite{web:selenuim} introduces a feature called Explicit Wait, which instructs the testing driver to wait for a specified amount of time until the presence of widgets.
Similar to Selenium, many tools incorporate this feature for mobile testing, such as Appium~\cite{web:appium}, UIAutomator~\cite{web:uiautomator}, etc.
Specifically, these tools verify the presence of widgets by fetching the view hierarchy of the GUI.
However, subsequent studies~\cite{li2020mapping} find that the fetched GUI views may be out of sync, leading to events on misaligned or invalid objects. 
While some studies~\cite{chen2019gallery,feng2022gallery,xie2020uied,xie2022psychologically,feng2021auto,feng2022auto,chen2020lost,feng2023video2action,chen2023unveiling} conduct UI modeling based on view hierarchy, they only check the validity of the GUI view hierarchy, not resource loading, which limits the effectiveness of replaying.
In contrast, we leverage the GUI as a whole with visual information to dynamically adjust the waiting time and schedule events when the GUI is fully rendered, which is analogous to human viewing and interaction.
Following the line of previous work~\cite{feng2023efficiency}, we integrate the GUI rendering classification model into the industrial app WeChat, offering insight into its potential for recording and replaying usage scenarios.

\section{Conclusion}
Record-and-replay is essential for ensuring quality assurance in the industrial app, WeChat.
Despite the numerous record-and-replay tools available, the waiting time between replaying events is often overlooked.
A short waiting time may hinder the effectiveness of replaying, while a long waiting time may reduce efficiency.
To address this, we propose a practical record-and-replay tool \tool, that employs a lightweight image-based approach to adaptively adjust the waiting time based on GUI rendering inference.
Given the real-time streaming on GUI, \tool uses a deep learning model to infer the rendering state and adjust event scheduling, replaying events when the GUI is fully rendered.
Experiments demonstrate the performance of our approach in improving the efficiency and effectiveness of recording and replaying within the WeChat app.

In the future, we plan to continue improving \tool's performance by collecting more GUI data from the WeChat app to enhance the accuracy of our model.
While \tool's motivation stems from WeChat, its adoption is not limited to this app, as it relies solely on GUI screenshots, which are easily obtainable in industrial apps.
We plan to explore its potential usage in other industrial apps and systematically evaluate its performance.

\bibliographystyle{ACM-Reference-Format}
\bibliography{main}

\end{document}